\documentclass[twocolumn,showpacs,preprintnumbers,amsmath,amssymb]{revtex4}
\usepackage{graphicx}
\begin{document}
\title{Factorizing Numbers with the Gauss Sum Technique: NMR
Implementations}
\author{T. S. Mahesh,
  Nageswaran Rajendran, Xinhua Peng, and Dieter Suter\footnote{Corresponding author:
Dieter Suter, dieter.suter@physik.uni-dortmund.de}}
\affiliation{Fachbereich Physik, Universit$\ddot{a}$t Dortmund, 44221 Dortmund, Germany\\
}
\date{\today}
\begin{abstract}

Several physics-based algorithms for factorizing large number were recently
published. A notable recent one by Schleich et al. uses Gauss sums for distinguishing
between factors and non-factors.
We demonstrate two NMR techniques that evaluate Gauss sums and thus implement
their algorithm.
The first one is based on
differential excitation of a single spin magnetization by a cascade
of RF pulses.  The second method is based on spatial averaging
and selective refocusing of magnetization for Gauss sums corresponding
to factors.  All factors of 16637 and 52882363 are successfully obtained.
\end{abstract}

\maketitle

\section{Introduction}  

Given an integer, factorizing it or confirming it to be a prime,
is an important problem in network and security systems \cite{kob1994}.
Using classical computers with finite resources of memory, 
the factorization time  
increases exponentially with the size of the number. 
In 1994, Shor discovered a quantum algorithm for factorization that takes only
polynomial time \cite{shor1994}.
Shor's algorithm requires a data register that is large
enough to encode the number to be factored along
with some ancilla qubits.  
A practical demonstration
has been carried out by factorizing the integer '15' 
\cite{nmrshor2001}, using nuclear magnetic resonance.
However, quantum computers capable of implementing Shor's algorithm for
larger number have not been developed so far.

Another physics-based method for factorizing numbers has been proposed
recently by the group of Schleich \cite{MAGPS2006,MBHSFS1_2002,MBHSFS2_2002}.
The mathematical basis of the technique relies on the properties of Gauss sums \cite{MerkelGsumMathFou}.
For the physical implementation, it requires an ensemble of two-level 
systems,
which accumulate the individual terms.
The sum is obtained by measuring the expectation value as an ensemble average.
As specific systems for the implementation of this algorithm,
atomic systems driven by resonant lasers have been proposed
\cite{MerkelCrasser2006, Merkel1foton, Merkel2foton}, and an implementation by nuclear spins $1/2$
was demonstrated by Mehring et al. \cite{gsumMehring2006}.

Like in Mehring's paper, we also use nuclear spins driven by 
radio-frequency (RF) pulses, but we use two different approaches that
demonstrate the flexibility of the Gauss-sum technique and require
fewer RF pulses.

\section{Gauss sums by differential excitation}

\subsection {Principle}

The factorization scheme of Schleich et al. relies on sums of the form
\begin{equation}
\sum_{m=-\infty}^\infty e^{i\,2\pi m^2 a} ,
\end{equation}
which are known as Gauss sums,
\footnote{Instead of 2, other exponents could also be used.
However, as in the papers of Schleich et al. we will restrict ourselves to 2,which provides efficient randomization.}.  Clearly, the series adds to infinity if $a$ is integer, and to zero otherwise.

Schleich et al. used this property for their factorization scheme by evaluating
the truncated series
\begin{equation}
%A_N^M(j)=\frac{1}{M+1}\sum_{m=0}^M\exp\bigg(2\pi \imath m^2\frac{N}{j}\bigg),
A_N^M(j)=\frac{1}{M+1}\sum_{m=0}^M\exp[i \phi_m(j)],
\end{equation}
where $\phi_m(j) = (2 \pi m^2\frac{N}{j})$,
$N$ is the number to be factored, the integer $j$ is trial factor, 
and $M$ is a truncation number.
This truncated series adds up to unity if $N/j$ is an integer,
i.e. if $j$ is a factor of $N$.
In all other cases, the sum is a small number whose value depends
on the trunction number $M$:
\begin{displaymath}
A_N^M(j) = \left\{ \begin{array}{ll}
 1,& \textrm{if $N/j=\textrm{integer}$},\\
 \ll 1, & \textrm{otherwise}.
  \end{array} \right.
\end{displaymath}

Physical systems that can implement this scheme must be described
by complex numbers. 
As a first example, consider a pendulum with two degrees of freedom.
Its excitation can be described by an amplitude and a direction angle,
which can be represented by a single complex number.
The individual terms of the Gauss sum are realized by 
resonant momentum transfers to the pendulum, 
with the direction specified by the phase of the complex number.
If $N/j$ is integer, the momentum transfers all occur in the same direction
and therefore keep increasing the amplitude of the pendulum.
In all other cases, the direction will vary and the individual momenta
interfere destructively. 

For the experimental realization, we choose a different system,
which is easier to realize: an ensemble of spins $I=1/2$, which
is excited by radio-frequency pulses. 
The Hamiltonian describing the interaction with the RF field is
\begin{eqnarray}
\mathcal{H}^{eff}_m = \omega \{ I_x \cos \phi_m(j) + I_y \sin \phi_m(j)\},
\end{eqnarray}
where $I_{x,y}$ are the components of the spin angular momentum
operator $\bf{I}$, $\omega$ is the strength of the RF field, and the phase angles
$\phi_m(j)$ are equal to the phases of the corresponding
terms in the Gauss sum. 

If the RF field is applied for a duration $\tau$, it rotates the spins
by an angle $\theta = \omega \tau$ around an axis in the $xy$-plane,
which is oriented at an angle $\phi_m(j)$ from the $x$-direction.
In close analogy to the mechanical pendulum, the effect of the individual
rotations adds coherently and reaches a maximum if the rotation axes
are the same (i.e. if $j$ factors $N$). In all other cases, the orientation
of the rotation axis is essentially random and the small rotations
cancel on average.

Formally, we describe the effect of a single pulse by the propagator 
\begin{eqnarray}
U_m(j) = \exp(-i \mathcal{H}^{eff}_m \tau).
\end{eqnarray} 

\begin{figure}[h]
\includegraphics[angle=270, width=3in]{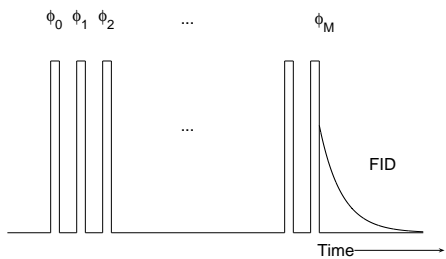}
\caption{
\label{method1ps}
Pulse sequence for the differential excitation method.}
\end{figure}

As shown in Figure \ref{method1ps}, the Gauss sum is evaluated by
applying a sequence of $M+1$ pulses, all with identical flip angle $\theta$
but variable phase $\phi_m$ to the spin system.
The combined effect of such a sequence of pulses is given by the propagator
\begin{eqnarray}
U(j) & = & U_M \cdots U_0  \nonumber \\
  & = & \prod_{m=M}^0 \exp[-i \theta \{ I_x cos\phi_m(j) + I_y sin\phi_m(j)\}],
\end{eqnarray}
where $\theta = \omega \tau$.
If $j$ is a factor, $\phi_m(j) = 2n\pi$ (with $n$ integer) and therefore
the pulses add coherently. 
The resulting net propagator is then
\begin{eqnarray}
U(j) = \exp[-i (M+1) \theta I_x].
\end{eqnarray}
As long as the total rotation angle $\theta (M+1)$ remains small compared
to $\pi/2$, the signal observed in the NMR spectrometer is proportional
to the angle and can be taken as a measure of the overall rotation. 
This implies that the individual rotation angles are small, $\theta \ll 1$.
Under this condition, the individual terms in equation (5) approximately
commute and the operator product can be evaluated also for non-factors.
Like in the mechanical analog, the individual contributions then
do not add coherently and therefore the effective 
transverse magnetization remains close to zero. 

\subsection {Experimental results}

In our experiments, we used the $^1$H spin of CHCl$_{3}$ dissolved
in Acetone-D$_6$ as our target system. 
The experiment was done on a 500 MHz Bruker Avance II+
spectrometer.  
Initially, the sample was in thermal equilibrium. 
The pulse sequence consisted of $(M+1)$ $[\theta]_{\phi_m}$ pulses
with the flip angle $\theta = 1^{\circ}$ and a delay of $5\mu s$
after each pulse, as shown in Figure \ref{method1ps}.
Since the reference frame is resonant with the Larmor precession of the spin, the free 
evolution operators during the delays become unit operators.  

At the end of the pulse sequence, we acquired the free induction decay signal,
which is proportional to the transverse magnetization.
For an accurate determination of its value, we calculated its Fourier 
transform and integrated over the resonance line.
The resulting data were normalized w.r.t. the 
reference spectrum (with $\phi_m = 0, \forall m $).

\begin{figure}[h]
\includegraphics[angle=270,width=3in]{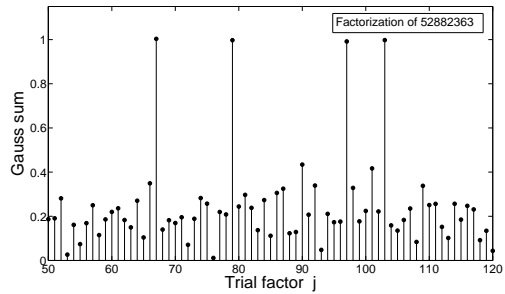}
\caption{
\label{method1result}
Factorization of 52882363 using the differential excitation method and 
the truncation number $M = 15$.}
\end{figure}

As a first example, we chose to factorize the
number $N = 52882363$.  The results are shown in Figure \ref{method1result}
 for trial factors $j$ between $50$ and $120$. 
It can be clearly seen that the factors
are 67, 79, 97, and 103.
The visibility of the data, i.e. the separation between the factors and nonfactors,
depends largely on the truncation number $M$.
For this example, we used a relatively small value of $M=15$.
For larger values, the relative separation can be increased significantly.

\section{Gauss sums by spatial averaging}

\begin{figure}[h]
\includegraphics[angle=270,width=3in]{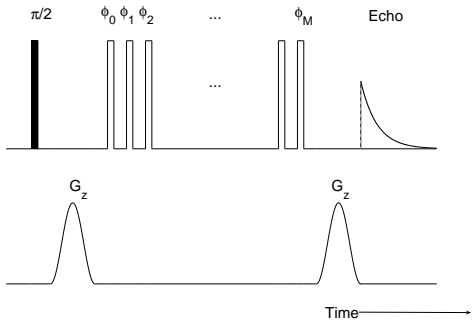}
\caption{\label{savmethod}
Pulse sequence for the spatial averaging method.
The initial pulse is an RF pulse that creates transverse magnetization.
The Gaussian-shaped second pulse is a magnetic field gradient pulse, 
and the subsequent rectangular pulses are RF pulses with flip-angle
$\theta$ and phase $\phi_m$.}
\end{figure}

%\subsection{Spatial averaging method}
Using the principles outlined above, many similar experiments can be
conceived that achieve the same objective. 
As a second example, we choose the case that is summarized in Figure \ref{savmethod}.
Here, we also encode the individual terms of the Gauss sum in the phase of
RF pulses.
In this case, we choose the flip angle $\theta$ of the individual pulses
such that the total flip angle adds up to
\begin{eqnarray}
(M+1)\theta = \pi
\end{eqnarray}
for factors. (More generally, it may be any angle $(2n+1)\pi$ with
n integer). 

This sequence of pulses is applied to a state where transverse magnetization
has been dephased by a magnetic field gradient pulse.
Starting from thermal equilibrium, where the density operator of the system
is proportional to $I_z$, the excitation pulse creates $I_x$-magnetization.
The field-gradient pulse turns this into
\begin{eqnarray}
I_x \stackrel{G_z}\longrightarrow & I_x \cos{\alpha_z}+ I_y\sin{\alpha_z} 
\end{eqnarray}
where $\alpha_z$ is the position($z$)-dependent dephasing introduced by 
the gradient,
\begin{eqnarray}
\alpha_z = \gamma G_z z T ,
\end{eqnarray}
with $\gamma$ the gyromagnetic ratio, $G_z$ the field gradient, $z$ the
coordinate in the direction of the gradient, and $T$ the duration of the gradient pulse.
The total signal, which is the integral of $I_x$ over space, vanishes at this point.
For factors $j$, the combined effect of the pulse sequence is a $\pi$-rotation,
which inverts the accumulated phase, independent of the position $z$
\cite{HahnEcho1950}:
\begin{eqnarray}
 I_x \cos{\alpha_z}+ I_y\sin{\alpha_z}   & \stackrel{\pi_x}\longrightarrow 
I_x\cos{\alpha_z}-I_y \sin{\alpha_z} .
\end{eqnarray}
When the second field gradient pulse acts on this state,
it adds another phase $\alpha_z$, which is identical to the first.
The final state is thus
\begin{eqnarray}
I_x\cos{\alpha_z}-I_y \sin{\alpha_z} &\stackrel{G_z}\longrightarrow &
(I_x\cos\alpha_z+I_y\sin\alpha_z) \cos\alpha_z \nonumber\\
&&-(I_y\cos\alpha_z-I_x\sin
\alpha_z) \sin\alpha_z  \nonumber \\
&=&I_x ,
\end{eqnarray}
i.e. the magnetization gets back into phase and an echo is observed.
This holds only for terms where $j$ is a factor of $N$; for the others,
the total effect of the sequence of $M+1$ pulses is not a phase reversal,
but only a small rotation, which cannot refocus the magnetization.

\begin{figure}[h]
\includegraphics[angle=270,width=3in]{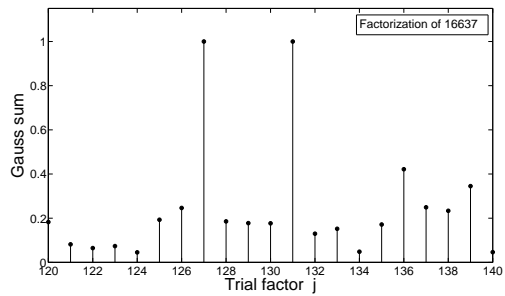}
\caption{\label{savresults}
Factorization of 16637 using spatial averaging method
with the truncation number $M = 12$.}
\end{figure}

In the experimental implementation, we used this scheme to determine the
factors of the number 16637.
Figure \ref{savresults} shows the experimental results for the trial factors $j$
between 120 and 140, when the sequence is truncated at $M = 12$.
Again, we can clearly distinguish between the factors (127 and 131) from
the non-factors.

\section{Conclusion}

Interference between different wavepackets is an important feature
of quantum mechanics. Schleich et al. have used this property to evaluate Gauss sums and shown how this can be used to determine if a given number is a factor of another number.  Their procedure is related to the proposal of Clauser and Dowling \cite{DowlingFactorization1996} who used optical interferometry for factorization.

In this paper, we discuss specific physical systems that implement such a summation
and give experimental examples that determine the factors of a given number.
In particular, we show two NMR experiments that calculate Gauss sums
and apply them to find the factors of numbers with 5 and 8 decimal digits.

The two techniques differ with respect to the conditions on the flip angles
as well as with respect to the required initial condition. 
We found that the visibility of the resulting scans is quite high,
even for small truncation numbers $M \sim 12 - 15$.  For larger numbers,
it might be necessary to use larger values of $M$ to obtain sufficient
contrast between the factors and non-factors.  

The spin system that we use for the implementation can be completely specified
in terms of a 2-dimensional Hilbert space.
Accordingly, the dynamics of the system can also be described in classical terms
and the algorithm may not be considered a quantum algorithm.
This fact is also evidenced by the example that we discussed (a classical pendulum).
In contrast to Shor's algorithm, this algorithm is no more efficient than known
classical algorithms.

\section{Acknowledgements}

This work was initiated by W. Schleich and M. Mehring, who also communicated 
their papers to us before publication.  We thank Jingfu Zhang for his
help in setting up the spectrometer and for useful discussions.
We gratefully acknowledge financial support from the Alexander von Humboldt Foundation, GK-726 Materials and Quantum Information Processing and the DFG through grant numbers Su192/19-1 and Su192/11-1.

\bibliography{gauss.bib}
\end{document}